\documentclass{article}
\begin{document}
\rightline{\vbox{\baselineskip=12pt{\hbox{CALT-68-2113}\hbox{QUIC-97-031}
\hbox{quant-ph/9705032}}}}

\bigskip
\centerline{\Large \bf Quantum Computing: Pro and Con}
\bigskip\bigskip
\centerline{\large John Preskill\footnote{\tt
preskill@theory.caltech.edu}}
\medskip
\centerline{\it California Institute of Technology, Pasadena, CA 91125, USA}
\bigskip
\begin{abstract}
I assess the potential of quantum computation.  Broad and important
applications must be found to justify construction of a quantum computer; I
review some of the known quantum algorithms and  consider the prospects for
finding  new ones. Quantum computers are notoriously susceptible to making
errors; I discuss recently developed fault-tolerant procedures that enable a
quantum computer  with noisy gates to perform reliably.  Quantum computing
hardware is still in its infancy;  I comment on the specifications that should
be met by future hardware.  Over the past few years, work on quantum
computation has erected a new classification of computational complexity, has
generated profound insights into the nature of decoherence, and  has stimulated
the formulation of new techniques in high-precision experimental physics.  A
broad interdisciplinary effort will be needed if quantum computers are to
fulfill their destiny as the world's fastest computing devices.

\smallskip
This paper is an expanded version of remarks that were prepared for a panel
discussion at the ITP Conference on Quantum Coherence and Decoherence, 17
December 1996.
\end{abstract}

\parskip=5pt 
\section{Introduction}

The purpose of this panel discussion is to explore the future prospects for
quantum computation.  It seems to me that there are three main questions about
quantum computers that we should try to address here:

\begin{itemize}
\item {\bf Do we {\it want} to build one?}  What will a quantum computer be
good for? There is no doubt that constructing a working quantum computer will
be a great technical challenge; we will be persuaded to try to meet that
challenge only if the potential payoff is correspondingly great.  To assess the
future viability of quantum computers,  we should thus try to imagine how they
will be used in the future.  Our imaginations are limited, so we are bound to
miss the most interesting applications; still, we should try.

\item {\bf {\it Can} we build one?} No one doubts that building a useful
quantum computer (one capable of addressing challenging computational problems)
is difficult, but is it actually {\it impossible}?  Do fundamental physical
principles pose a truly serious obstacle?  The most obvious concern is the
problem of errors.  As for an analog classical computer, the errors made by a
quantum computer form a continuum, and so the accumulation of small errors
threatens to destabilize the device. Furthermore, quantum computers rely for
their special capabilities  on quantum entanglement (non-classical correlations
involving many degrees of freedom), and entanglement is particularly vulnerable
to the effects of decoherence due to uncontrollable interactions with the
environment.  Can these difficulties be overcome even in principle (and if so,
also in practice)?

\item {\bf {\it How} will we build one?} What kind of hardware will the quantum
computers of the future use?  Can this hardware be constructed via incremental
improvements of existing technologies, or will truly new ideas be needed?
\end{itemize}

\noindent I do not know the answers to these questions, but I will express a
few thoughts.  

\section{Do we {\it want} to build one?}

How will quantum computers be used?

I have enormous admiration for Peter Shor's factoring algorithm (Shor 1994).
With Shor's algorithm, it is possible to find the prime factors of an $N$-digit
number in a time of order $N^3$, while it is widely believed (though it has
never been proved) that any factoring algorithm  that runs on a classical
computer would require a time that increases with $N$ faster than any power.
This stunning result, and the ingenuity of the algorithm, aroused much interest
in quantum computation.\footnote{Daniel Simon (1994) actually paved the way for
Shor's factoring algorithm, by exhibiting the first example of a quantum
algorithm that efficiently solves an interesting hard problem.} The desire for
a powerful factoring engine (with cryptographic applications) is now widely
regarded as one of the primary motivations for building a quantum computer.
But in the long run, I do not expect factoring to be one of the most important
applications of quantum computing. In fact it seems to me that it is largely a
historical accident that the factoring problem is regarded as especially
important today.

If not factoring, what then?  I am actually quite sympathetic with Feynman's
original vision (Feynman 1982) -- that a quantum computer will be used to
simulate the behavior of quantum systems.\footnote{However, my  view that
quantum simulation is more important than factoring was dismissed by some
participants at the conference as the bias of a narrow-minded physicist --- one
who thinks that the only important problems are the ones that he works on!}
What Feynman emphasized is that a quantum device can store quantum information
far more efficiently than any classical device; since $N$ qubits live in a
Hilbert space of dimension $2^N$, a classical device would record $2^N-1$
complex numbers to describe a typical quantum state of the $N$ qubits. So it is
plausible that quantum simulation is an example of a task that requires
exponential resources for a classical computer, but not for a quantum
computer.\footnote{But it was David Deutsch (1985), not Feynman, who emphasized
that quantum computers can best realize their computational potential by
exploiting massive quantum parallelism.  That a quantum system can perform
computation was first explicitly pointed out by Benioff (1982).} (Exponential
memory space is not really necessary, but presumably the simulation requires
exponential time.) Furthermore, quantum simulation is a rich subject, with many
potential applications to, say, materials science and chemistry.  I think it is
important to flesh out in more detail how quantum computers would be used as
quantum simulators, so that we can better assess the advantages that  quantum
computers will enjoy over future classical computers (Lloyd 1996; Zalka 1996a; Wiesner 1996;
Meyer 1996; Lidar \& Biham 1996; Abrams \& Lloyd 1997; Boghosian \& Taylor 1997). In principle, a
question about a quantum system becomes exceptionally difficult only when the
answer depends delicately on the details of entanglement involving many degrees
of freedom, and it is not clear for what physically interesting questions
massive entanglement plays an essential role.   For example, few-body
correlations in the ground state can usually be computed in polynomial time on
a classical computer.  Classical simulations of the real-time evolution of a
quantum system seem to be more challenging, but perhaps, with sufficient
ingenuity, new approximations can be developed that will vastly improve the
efficiency of such simulations.  Using a quantum computer is more of a brute
force strategy; yet sometimes we save much effort by invoking the brute force
method rather than the one that requires exceptional cleverness. 

Since Shor exhibited the factoring algorithm, perhaps the most important new
development in quantum complexity has been Grover's very clever method for
searching an unsorted database (Grover 1996).  In a database containing $N$
items, the one item that meets a specified criterion can be found with a
quantum computer in a time of order $\sqrt{N}$.   On a classical computer, the
database search would take a time of order $N$, so Grover's algorithm is a case
where we {\it know} that the quantum computer can perform a computationally
important task faster than any classical computer.  (This hasn't been proved
for factoring, though it is likely to be true in that case, too.) The speedup
is achieved by exploiting both quantum parallelism and the property that a probability in quantum theory is the square of an amplitude --- Grover's algorithm acts on an
initial state in which all of $N$ classical strings are each represented with
an amplitude $1/\sqrt{N}$, and rotates the initial state in of order $\sqrt{N}$
steps to a state in which the string being sought is represented with an
amplitude of order one.

The speedup relative to classical methods achieved by Grover's algorithm is not
nearly so spectacular as the exponential speedup achieved by Shor's algorithm.
But even a non-exponential speedup can be very useful.  The database search is
surely an important problem with many applications; for example, it could be
used to solve any NP problem (a problem whose solution, though perhaps hard to
find, is easy to verify). If quantum computers are being used 100 years from
now, I would guess that they will be used to run Grover's algorithm or
something like it. Furthermore, it seems likely that there is much more to say
about algorithms like Grover's that provide a non-exponential speedup.  In the
case of the database search, a computation that requires time $T$ on a
classical computer can be performed in a time of order $\sqrt{T}$ on a quantum
computer.  It would be quite interesting to find a general way to characterize
the classical algorithms that will admit this kind of $\sqrt{T}$ quantum
speedup.  In particular, classical computers usually address NP-complete
problems not by doing a blind search for the desired solution, but by doing a
search that is considerably more clever and more efficient, and that still
succeeds reasonably well.  To what extent can these more efficient algorithms
be improved by means of quantum computation?  

Speculation about the prospects for quantum computing often centers on the
issue of NP-complete problems, and especially the dream that quantum
computation will allow an exponential speedup of the solution to problems in
this class. In this connection, an important result was obtained by Bennett,
Bernstein, Brassard, and Vazirani (1997a), who showed that Grover's algorithm
for the database search problem is actually optimal; no quantum algorithm can
solve the problem faster than time of order $\sqrt{N}$.  This result suggests
that quantum computers may not prove capable of solving NP-complete problems in
polynomial time.  At the very least it indicates that no such polynomial-time
quantum algorithm will exist that relies on sheer ``quantum magic;''  rather,
penetrating insights into the structure of the problems in the NP-complete
class may well be required.  

Perhaps, though,  NP-complete problems are not the best context for exploiting
the power of quantum computation. 
It may be that quantum computers are capable of solving some hard problems that
are {\it outside} NP, and that quantum simulation is an example.  Even were an
oracle at our command that could solve NP-complete problems, quantum simulation
might still require exponential resources on a classical computer; that is, a
classical computer would still be unable to simulate a quantum computer
efficiently.\footnote{In fact, a weakened version of this statement (``relative
to an oracle'') was demonstrated by Bernstein and Vazirani (1993).} Quantum
computing is likely to have the most dramatic payoff in algorithms that best
exploit the ability of a quantum storage register to store an exponentially
complex quantum state using polynomial quantum resources.  (And for these
algorithms, it is bound to be crucial that the quantum computer can generate
highly entangled quantum states.)

Assuming the correctness of the central conjecture of classical complexity
theory (${\rm }P\ne {\rm NP}$), there exists a class of problems (the class
NPI) of intermediate difficult; these problems are not as hard as the
NP-complete problems, yet still cannot be solved by a Turing machine in
polynomially bounded time.  The factoring problem is regarded as a likely
candidate for membership in this class (Garey \& Johnson 1979), and so it is
natural to wonder whether efficient quantum algorithms can be devised for other
problems that are suspected to be in NPI.  One particularly promising example
is the graph isomorphism problem (to determine whether two specified graphs are
equivalent after a suitable relabeling of the vertices).  It is important to
investigate whether good quantum algorithms can be devised for the graph
isomorphism problem and other related problems.

I feel that a deep understanding of {\it why} quantum algorithms work is still
lacking.  Surely the power of quantum computers has something to do with
entanglement, quantum parallelism, and the vastness of Hilbert space, but I
think that it should be possible to pinpoint more precisely the true essence of
the matter.  One form of the question is:  how does Planck's constant $\hbar$
enter into quantum computation, and what is the nature of the ``classical''
limit $\hbar\to 0$?  I suspect that a better understanding of this sort of
issue would help to point us toward new types of quantum algorithms.

In another talk at this conference (Preskill 1997), I estimated the resources
that would be needed to solve an interesting factoring problem on a quantum
computer.  The estimate was surely daunting.  Perhaps an interesting quantum
simulation problem could be effectively addressed with more modest resources.
But it is also natural to ask what could be done with a small quantum computer,
one that can store, say, tens of qubits and can implement hundreds of gates.
If we could build such a device reasonably soon, would it be useful?  Would it
have commercial potential?

One possible application of a quantum computer of modest size would be to
quantum cryptography (Bennett \& Brassard 1984).  Of course, in the absence of
quantum factoring engines, conventional public key cryptography may be secure,
but I assume there will always be some users who will insist on the ultimate in
privacy, and so will prefer quantum key distribution.  (For one thing, the user
might fear that his message could be stored and deciphered some time in the
future, when more powerful factoring techniques become available.) Though
quantum key distribution may be secure in principle, it has a serious
limitation:  the signal becomes attenuated in the communication channel (such
as an optical fiber), and {\it it cannot be amplified}, because of the
no-cloning theorem (Wootters \& Zurek 1982).  So either we must be satisfied
with communication that is limited to distances of the order of the attenuation
length in the fiber (perhaps tens of kilometers), or we must be willing to
enlist trusted intermediaries, which would clearly entail a serious security
risk.  But quantum error correction may provide an alternative: if we could
prepare, send, and receive  {\it entangled} multi-photon states, then in
principle we could use quantum error-correcting codes to extend the range of
quantum communication.  ``Repeaters'' would be placed along the communication
line; these repeaters would not read the quantum information that is being
transmitted; rather they would diagnose and correct the errors that occur
during transmission.  We could send, say, blocks of five photons that encode
one logical qubit (Bennett {\it et al.} 1996; Laflamme {\it et al.} 1996)
chosen at random to be in one of two non-orthogonal states, and place the
repeating stations close enough together that the probability of error during
transmission between successive stations is small.  Our quantum computers would
need to be capable of carrying out (fault-tolerant) syndrome measurement and
error correction for the five-qubit code with a small probability of error
(Shor 1996; DiVincenzo \& Shor 1996), and we would need to be able to quickly
refresh the ancilla bits that are used to compute the syndrome in order to achieve a reasonable transmission rate. Of course, with more powerful quantum computers, we could use better codes and improve the performance of the network.

Perhaps the best clocks of the reasonably near future will have quantum
computers inside. Some atomic clocks are limited by the spontaneous decay
lifetimes of the excited atomic states.  If error-correcting codes could be
used to inhibit spontaneous decay, then in principle longer interrogation times
could be achieved and a more precise frequency standard could be established.
The NIST group (Bollinger {\it et al.} 1996) has suggested another way that
quantum entanglement could be invoked to improve the precision of a clock or an
interferometer. If the phase oscillations of the state
${1\over\sqrt{2}}\left(|0\rangle + |1\rangle\right)$ of a two-level system are
used to define a frequency standard, then the ``cat state''
${1\over\sqrt{2}}\left(|000\dots 0\rangle + |111\dots 1\rangle\right)$
constructed from $N$ such systems would oscillate $N$ times as fast, and so
could in principle be used to establish a more precise standard.\footnote{However, the improvement in precision that can be attained by using entangled states will be severely limited by decoherence effects --- while entangled states oscillate faster than unentangled states, they also decohere more rapidly (Huelga {\it et al.} 1997).} 
  
Even if the commercial potential of a low-performance quantum computer might be
modest, it could well become an essential tool in the laboratory of the
experimental physicist.  The ability to prepare, maintain, manipulate, and
monitor highly entangled states will make it possible to perform a wide variety
of ingenious new measurements.  

But suppose I could buy a truly powerful quantum computer off the shelf today
--- what would I do with it?  I don't know, but it appears that I will have
plenty of time to think about it!  My gut feeling is that if powerful quantum
computers were available, we would somehow think of many clever ways to use
them.

\section{{\it Can} we build one?}

Manny Knill and I have talked at this meeting about the remarkable recent
progress in the theory of fault-tolerant quantum computation.  Even before the
recent developments, one might have been hopeful that error-correction methods
could be invoked to resist decoherence and to control the accumulation and
propagation of error in a quantum computer, but this view could have been
dismissed as wishful thinking. Now the case is considerably stronger that, in
principle, there is no fundamental obstacle to building a functioning quantum
computer that is capable of performing computationally interesting tasks.

Serge Haroche and Rolf Landauer have argued eloquently at this meeting that
this point of view is much too sanguine, and perhaps it is.  Haroche argues
that the optimists grossly underestimate the pervasiveness of decoherence and
the difficulty of resisting it (Haroche 1997; Haroche \& Ramond 1996). Haroche
notes that a highly entangled state of many qubits is exceptionally vulnerable
to the effects of decoherence --- a single error affecting just one of the
qubits can destroy the coherence of the whole state.  Indeed, this is so; in a
functional quantum computer, quantum error correction must work so effectively
that hardly a single logical (encoded) qubit fails during the course of the
computation.  He also emphasizes that error-correcting codes entail an enormous
overhead in a quantum computation, both in terms of the number of qubits
required (to provide the necessary redundancy to recover from errors) and the
number of quantum gates needed (to process the redundantly encoded data, and to
diagnose and reverse the errors); this increase in the number of qubits and the
number of gates increases the likelihood of error. Indeed this is so, as I have
discussed in my other talk at this conference (Preskill 1997).  But it has now
been shown that if the error probability per gate is less than a certain
critical value (the ``accuracy threshold''), then error correction can still
work effectively, even for a computation that is arbitrarily long (Knill \&
Laflamme 1996; Knill {\it et al.} 1996, 1997; Aharonov \& Ben-Or 1996a; Kitaev
1996b; Gottesman {\it et al.} 1996; Zalka 1996b; Preskill 1997).

If fault-tolerant methods are invoked to improve the reliability of a quantum
computer, then a price must be paid in both storage requirements and processing
speed.  However, this price may be quite acceptable. Given hardware with a
fixed rate of error per elementary gate, to do a longer computation with
acceptable accuracy, we will need to increase the block size of the code.  But
the scaling of the needed block size with the length of the computation to be
performed is reasonably favorable:
\begin{equation}
{\rm Block ~ Size}\sim \left[\log\left({\rm Length~ of~
Computation}\right)\right]^{\rm power}\ .
\end{equation}
(In the scheme described in Preskill (1997), the power is $\log_2 7\simeq
2.8$.) To process the information encoded in these larger blocks also requires
more elementary gates (with the number scaling roughly linearly with block
size).  However, in principle, many of these gates could be performed {\it in
parallel}. {\it If} we assume that the quantum hardware is highly
parallelizable, then the processing time is only weakly dependent on the block
size.

It has been suggested that fault-tolerant procedures will not deal effectively
with errors in which qubits ``leak'' out of the Hilbert space on which the
quantum computer acts (Plenio \& Knight 1996).  For example, in an ion trap one
of the ions might make an unwanted transition to a long-lived inert state that
is not acted on by the quantum gates of the machine.  Such errors will
inevitably occur, but they need not pose a serious obstacle. One possible
strategy is to systematically pump the levels that are prime candidates for
leakage.  But furthermore, leakage errors can easily be detected in principle
with a simple quantum gate array (Preskill 1997).  An ion identified as faulty
can be eliminated from the code block and replaced by a standard ion in the
ground state.  After the replacement, a leakage error becomes an error in a
known location that can easily be dealt with using standard error-correction
procedures (Grassl {\it et al.} 1996).

Haroche also questions whether it will be possible to achieve error rates per
gate that are small enough for quantum computers to work accurately.  The
fundamental difficulty is that qubits must interact strongly if we are to
fashion a quantum gate; but establishing this strong interaction may also
induce the qubits to interact with other degrees of freedom (the environment),
which will lead to decoherence.  For example, to improve the performance of an
ion-trap computer, we would increase the laser intensity to speed up the gates.
But as the intensity increases, it becomes more likely that the ion is excited
to a different level than was intended.  The competition between these two
effects imposes an intrinsic limit on the accuracy of the gate that is
insensitive to the choice of the ion used (Plenio \& Knight 1996). Under fairly
general assumptions, one concludes that the probability of error per gate is at
least of order $10^{-6}$. This limit might be evaded through suitably heroic
efforts -- for example by placing ions in small cavities that are carefully
engineered to suppress the unwanted transitions. Nevertheless, arguments of
this sort are unquestionably useful, and it would be of great interest to
formulate general limits that would constrain other types of transitions or
other conceivable hardware implementations.\footnote{A very weak general
limitation on the performance of quantum hardware due to the vacuum
fluctuations of the electromagnetic field was pointed out by Braginsky,
Khalili, and Sazhin (1995).  In the context of an ion trap, their limit arises
because the phonon in the trap can in principle decay by emission of a {\it
photon}. (See also Garg (1996).)} 

Even if it proves difficult to improve on an error rate of order $10^{-6}$ per
gate, hardware that approaches this level of accuracy may already be suitable
for reliable large-scale quantum computation. Indeed, in my talk on
fault-tolerant quantum computation at this conference (Preskill 1997), I
suggested that an error rate of $10^{-6}$ per gate is a reasonable target to
aim for --- this error rate is likely to be sufficiently below the accuracy
threshold that very interesting quantum computations will be feasible with
reasonable resources.\footnote{Any statement about acceptable error rates is
meaningless unless a model for the errors is carefully specified.  In Preskill
(1997), uncorrelated stochastic errors are assumed. Under this assumption, all
errors are due to decoherence, phase errors and bit flip errors are equally
likely, and errors affecting distinct qubits are independent. The ``gate error
probability'' $\epsilon\sim 10^{-6}$ may be interpreted as a quantum fidelity
-- that is, if the computer would have been in the state $|\psi\rangle$ had the
gate been implemented perfectly, and if its actual state after the gate is
applied is $\rho$, then $F\equiv\langle \psi |\rho | \psi\rangle =1-\epsilon$.}
But I don't want to give the impression that this accuracy requirement is
etched in stone; it may be too conservative for a number of reasons.  First of
all, this estimate was obtained under the assumption that phase and amplitude
errors in the qubits are equally likely.  With a more realistic error model
better representing the error probabilities in an actual device, the error
correction scheme could be better tailored to the error model, and a higher
error rate could be tolerated.  Furthermore, even under the assumptions stated,
the fault-tolerant scheme has not been definitively analyzed; with a more
refined analysis, one can expect to find a somewhat higher accuracy threshold,
perhaps considerably higher.  Substantial improvements might also be attained
by modifying the fault-tolerant scheme, either by finding a more efficient way
to implement a universal set of fault-tolerant gates, or by finding a more
efficient means of carrying out the measurement of the error syndrome.  With
various improvements, I would not be surprised to find that a quantum computer
could work effectively  with a probability of error per gate, say, of order
$10^{-4}$. (That is, $10^{-4}$ may be comfortably {\it below} the accuracy
threshold. In fact, estimates of the accuracy threshold that are more
optimistic than mine have been put forward by Zalka (1996b). See also Steane
(1997).)  Another point that should perhaps be emphasized is that as the error
rates improve, it becomes possible to make more efficient use of storage space,
by using codes that encode many logical qubits in a single block.  Gottesman
(1997) has recently shown how to carry out fault-tolerant computation using
such codes, though at a cost in processing time.

Rolf Landauer has played a valuable role, in his remarks at this meeting and in
his previous writings (Landauer 1995, 1996, 1997), by reminding us that
proposed new technologies rarely realize the rosy projections put forth by
their proponents.  And he has repeatedly stressed the crucial issue of error
control (see also Unruh (1995).  Landauer correctly points out that digital
devices can achieve remarkable reliability because a digital signal can be
easily restandardized --- that is, if it wanders slightly from it's intended
value, it can be shoved back where it belongs.  This restandardization, which
prevents small errors from accumulating and eventually becoming large errors,
is necessarily a {\it dissipative} process.  Ease of restandardization is the
central advantage that digital devices enjoy over analog devices.  Quantum
computation (or more generally reversible computation) may seem from this point
of view to be an ill-devised return to analog computation, with all the
attendant problems.

It has been a truly stunning discovery that, using quantum error correction, it
is actually possible to restandardize a coherent quantum signal (Shor 1995;
Steane 1996ab; Calderbank \& Shor 1996).  Of course, like any error-correction
technique, quantum error correction is a dissipative process and so produces
waste heat that must escape from the device.  In a quantum error-correction
scheme, information about the errors that have occurred accumulates in a set of
ancilla qubits.  If these ancilla bits are to be reused, they must first be
cleared, which means that the entropy associated with the accumulated errors
must be released to the environment.  This need for cooling to remove the
entropy associated with the errors may be an important engineering constraint
on the quantum computers of the future.

Landauer acknowledges that significant progress has been made on the problem of
error control, but he also raises some vexing questions.  A question he has
raised insistently at this meeting is: how can coding prevent our quantum gates
from making small errors, if the code and the device have no way of knowing
what gate we are trying to implement? If we make a small mistake while
performing a gate that acts on encoded qubits, the final state of the qubits
{\it might} still reside in the code subspace, but take a value that differs
slightly from what we intended.  This kind of error is undetectable and
uncorrectable.  Won't such small errors inevitably accumulate over time and
result in large errors?  (The concern being raised has little to do with
decoherence; even if the evolution of the state of the computer is unitary,
there is no {\it a priori} guarantee that the unitary evolution is as
desired.)

We can assess this objection for the case of the trivial gate, where the
unitary transformation that we intend to apply is the identity.\footnote{The
reasoning can easily be adapted to the case of a nontrivial gate.}  Imagine
that transformations that differ slightly from the identity are applied to each
of the elementary qubits in the block, so that if the qubits are subsequently
measured, the probability of a phase or bit flip error in each qubit is of
order $\epsilon<<1$.  We may ask about the probability of an undetectable error
-- with what likelihood does the block still lie in the code subspace, but with
the encoded qubit pointing in the wrong direction?  For a code that is capable
of recovering from a single one-qubit error at an arbitrary position in the
block, this probability is actually of order $\epsilon^3$;  three independent
phase or bit flip errors are required for the {\it block}, when measured, to
take a value in the code subspace that differs by a phase or bit flip error
from it's original value.\footnote{It is actually more likely, occurring with
probability of order $\epsilon^2$, that upon performing fault-tolerant error
correction we misdiagnose the error in the block and reset the qubit to an
incorrect value.} 

Regarding nontrivial gates, it is important to note that the fault-tolerant
operations that can be performed on (say) a single encoded qubit do not form a
continuum; instead, only a discrete set of transformations can be safely
implemented.  Thus, small errors in the gate implementation, rather than
changing the intended gate to a different gate in the fault-tolerant set, will
be much more likely to cause detectable errors that can be corrected. Of
course, even though the set of fault-tolerant gates is discrete, it may still
be {\it universal}; if we have a universal set of fault-tolerant gates, we can
surely use them to construct a transformation that comes arbitrarily close to a single-qubit rotation with any desired angle, but we will need to use some of our multi-qubit gates in
that construction.

Landauer also reminds us that the efficacy of error correction will be reduced
if the errors have a systematic component.   Errors with random phases
accumulate like a random walk, so that the {\it probability} of error
accumulates roughly linearly with the number of gates applied.  But if the
errors have systematic phases, then the error {\it amplitude} can increase
linearly with the number of gates, and the probability of error might become
appreciable much sooner.  Hence, for our quantum computer to perform well,  the
rate of systematic errors must meet a more stringent requirement than the rate
for random errors.  Crudely speaking, {\it if} we assume that the systematic
phases always conspire to add constructively, and if the accuracy threshold is
$\epsilon$ in the case of random errors, then the accuracy threshold will be
approximately $\epsilon^2$ in the case of systematic errors; of the order of
$10^{-10}$, say, instead of order $10^{-5}$.  While systematic errors may thus
pose a challenge to the quantum engineers of the future, they ought not to pose
an insuperable obstacle. First, systematic phases will tend to cancel out over
the course of a long computation, so that higher error rates could be tolerated
in practice (Obenland \& Despain 1996, 1997; Miquel {\it et al.} 1997).  And
furthermore, if errors really are systematic, we can in principle understand
their origin and eliminate them.   It is always the random errors that place
the intrinsic limitations on performance.  

There is another important respect in which the error models that have been
used in theoretical studies of fault-tolerant quantum computing may be
unrealistic -- it is typically assumed that the errors afflicting distinct
qubits are {\it uncorrelated} or only weakly correlated.  In fact, this is a
very strong assumption, and an essential one, because quantum error-correcting
codes are not equipped to deal with strongly correlated errors involving many
qubits.  When we say that the probability of error  is of order $\epsilon\sim
10^{-6}$ per gate, we actually mean that the probability of two errors
occurring in a single block is of order $\epsilon^2\sim 10^{-12}$.  It will
ultimately be an experimental question whether different qubits in the same
block can really be decoupled to this degree.  We should note, though, that
there is no reason why two qubits belonging to the same code block need to be
near each other in the machine.  Thus, we have the opportunity to enhance the
validity of our error model through a suitable choice of machine architecture.

Now that we are convinced that quantum-error correction is possible, we should
search for new ways to implement it.  It is  important in particular to analyze
in greater detail how error correction methods can be adapted to some of the
proposed realizations of quantum hardware, in particular to ion-trap and
cavity-QED computers (Pellizzari {\it et al.} 1996; Mabuchi \& Zoller 1996; Van
Enk {\it et al.} 1997).
Furthermore, while most of the work on quantum error correction schemes has
focused on the quantum circuit model --- suitable circuits are designed to
diagnose and correct the errors --- a broader viewpoint might be highly
productive. One alternative procedure would be to devise a ``designer
Hamiltonian'' which has the protected code subspace as its (highly degenerate)
ground state.\footnote{One might hope that the designer Hamiltonian could be
realizable in a suitable mesoscopic implementation of quantum computation.}
Then an error would typically occur only when the system makes a transition to
an excited state, and this error would be automatically corrected when the
system relaxed to the ground state. Schemes of this sort have been
suggested by Kitaev (1996a).  Kitaev (1997) has also made the ingenious suggestion that
fault-tolerant quantum gates might be realized in a suitable medium by
exchanging quasiparticles that obey an exotic version of two-dimensional
quantum statistics.  The essential idea is that, because the long-range
Aharonov-Bohm interactions among the quasiparticles would be {\it topological},
the gate would not have to be implemented with high precision in order to act
in the prescribed way on the quantum numbers of the quasiparticles.

Landauer urges us to consider that, even if a quantum computer can be
constructed, and even if it is capable of performing highly valuable tasks, the
technology will have little impact if it is absurdly expensive.  Again, this is
a serious objection.  Surely, the technology has far to go before we can even
begin to seriously assess the economics of quantum computation.  The more
important point, though, is that to be economically viable, a quantum computer
would have to have broad applications.  Searching for (and finding) new and
useful quantum algorithms may be the most effective way of bringing quantum
computing closer to fruition as a commercial enterprise.

Some of the commonly expressed reasons for skepticism about quantum computing
are listed in Table \ref{tab:objections}, along with some countervailing
views.

\begin{table}
\caption{Some possible objections to quantum computation, and some responses}
\label{tab:objections}
\begin{center}
\begin{tabular}{|p{3.0in}|p{3.0in}|}
\hline
{\bf Objection} & {\bf Response}\\
\hline\hline
Quantum computers are analog devices, and hence cannot be restandardized. &
Using quantum error-correcting codes and fault-tolerant error correction, we
{\it can} restandardize encoded quantum information.\\
\hline
A quantum error-correcting code cannot detect or correct an error if the
quantum state remains inside the protected code subspace.  &  If good codes are
used, such uncorrectable errors are quite unlikely.\\
\hline
A high-performance quantum computer must prepare and maintain highly entangled
quantum states, and these states are very vulnerable to decoherence. &
Fault-tolerant error correction protects highly entangled encoded states from
decoherence.\\
\hline
Error correction itself requires a complex computation, and so is bound to
introduce more errors than it removes! & If the error probability per gate is
below the {\it accuracy threshold}, then an arbitrarily long computation can in
principle be performed with negligible probability of error.\\
\hline
Quantum error correction will slow down the computer. & With highly
parallelized operation, the slowdown need not be serious.\\
\hline
To successfully incorporate quantum error correction, a much larger quantum
computer will be needed. & The number of qubits needed increases only
polylogarithmically with the size of the computation to be performed. \\
\hline
Any quantum computer will suffer from leakage -- the quantum information will
diffuse out of the Hilbert space on which the computer acts. & With suitable
coding, leakage errors can be detected and corrected.\\
\hline
Systematic errors will accumulate over time; error-correcting codes do not deal
with systematic errors as effectively as with random errors. & In principle,
systematic errors can be understood and eliminated.\\
\hline
Coding does not protect against highly correlated errors. & Correlated errors
can be suppressed with suitable machine architecture.\\
\hline
There are intrinsic limitations on the accuracy of quantum gates.  Error
correction will not work for gates of feasible accuracy. & Within the known
limits, gates exceeding the accuracy threshold are possible. With suitable
hardware, even these limits might be evaded.\\
\hline
Current quantum computing technology is inaccurate, slow, not scalable, and not
easily parallelizable. & Faster gates and new ways to distribute entanglement
can surely be developed.  New ideas for quantum hardware would be most
welcome!\\
\hline
In the near term, experiments with quantum computers will be mere
demonstrations.  They will not teach us anything. & We will learn about
correlated decoherence.  The performance of devices with just a few tens of
qubits cannot be easily simulated or predicted.\\
\hline
Quantum computers will be too expensive. & But they will be worth the price if
suitably broad applications are found.\\
\hline
The known applications of quantum computers are quite limited. & Let's think of
new ones!  The known quantum database search algorithm may prove to be very
useful.\\
\hline
\end{tabular}
\end{center}
\end{table}

\section{{\it How} will we build one?}

Quantum computing hardware is clearly in its infancy.  Though what has already
been achieved using ion traps (Monroe {\it et al.} 1995), cavity QED (Turchette
{\it et al.} 1995), and NMR techniques (Cory {\it et al.} 1996; Gershenfeld \&
Chuang 1997) is intriguing and impressive, all of these technologies have
serious intrinsic limitations.  The quantum computing hardware of the future is
bound to be substantially different than the hardware of the present.

For example, the speed of an ion trap computer operated according to the
Cirac-Zoller (1995) scheme is limited by the frequencies of the vibrational
modes in the trap.  In the original NIST experiment (Monroe {\it et al.} 1995),
this frequency was about 10 MHz, but it is likely to be orders of magnitude
smaller in a trap that contains multiple ions.  NMR devices suffer from an
exponential attenuation of signal to noise as the number of qubits in the
machine increases.  Of the ``current'' quantum computing technologies, those
based on cavity QED probably have the best long-term potential (Cirac {\it et
al.} 1996).

Future hardware will have to be fast, scalable, and highly parallelizable.
Indeed, parallel operations will be crucial to error correction.  In addition
to errors introduced by the quantum gates themselves, we will also need to
worry about {\it storage} errors, those affecting the ``resting'' qubits that
are not acted on by the gates.  To control storage errors, it will be necessary
to perform error correction continually on the resting qubits, which will be
infeasible in a large device unless many code blocks can be corrected
simultaneously. Even apart from the issue of storage errors (and with
appropriate hardware design, they may not be a serious limitation), parallel
operation is highly desirable to improve processing speed.  This is especially
so because quantum error correction will slow down a computation substantially
unless it is possible to operate simultaneously on many qubits in the same code
block.

Thus it will be essential for our machine to be able to distribute the parts of
a highly entangled state to various processors, and to act on the parts
independently.  In an ion trap machine, for example, there would be many traps,
each containing multiple qubits, and the machine would have to be able to shunt
ions from one trap to another without disturbing the internal atomic states.
In the case of a cavity QED device, a promising suggestion for distributing
entanglement was made by Cirac, Zoller, Kimble, and Mabuchi (1997); in their
scheme, atoms are trapped in many cavities, and entanglement is established
among atoms in distinct cavities by exchanging photons between the cavities.

Scalability obviously is going to be a crucial issue if we ultimately hope to
build machines that are capable of storing and manipulating millions of
individually addressable qubits.  In the long term, it may be some sort of
solid state or microfabricated device that will have the most promise of
offering the needed scalability.  We should also be prepared to adapt our
paradigm for quantum computing to new technological opportunities.  For
example, as Lloyd (1993) has advocated, a molecular machine is likely to
operate more like a quantum cellular automaton than like the quantum gate
circuits that most theorists usually envision.  

Even more broadly, since no quantum system that is well protected from
decoherence can be efficiently simulated on a classical computer, any such
system has the potential to perform difficult computational tasks. Aharonov and
Ben-Or (1996b) have observed that a phase transition can be expected as the
decoherence rate is varied.\footnote{This is related, at least morally, to the
sort of phase transition in dissipative quantum systems discussed some time ago
by Leggett {\it et al.} (1987)}  A very noisy quantum system behaves
classically, and so can be efficiently simulated with a classical Turing
machine, but if the decoherence rate is low enough, highly entangled quantum
states can be established and no efficient classical simulation is possible.
In this sense, any quantum system with a low decoherence rate is performing a
hard ``computation.''  For a theorist familiar with critical phenomena, it is
natural to wonder about the universal characteristics of this phase transition
-- for example,  it would be interesting to compute the critical exponents that
govern the scaling properties of the transition, as these would not depend on
the particular microscopic Hamiltonian of the system being considered.  Insight
into this question might suggest new physical implementations of quantum
computation.

The gap between current quantum computing technology and what will be needed in
the future is so vast that one can easily be discouraged. But we should not
accept the (justified) criticism of the existing technology as a damning
assessment of the ultimate prospects.  Rather, mindful of the potential power
of the quantum computer, we should be energized by the challenge to fabricate
the hardware that will make it work.

\section{Quantum computing at the fin de si\`ecle}

Quantum computing may be the technology of the day {\it after} tomorrow.  But
most of us are not so patient.  
What can we and should we be doing tomorrow?  Or today?

It now seems likely that the first experiments to perform a computation
involving several quantum gates will be carried out using the NMR method (Cory
{\it et al.} 1996; Gershenfeld \& Chuang 1997), and will soon be followed by
experiments with ion trap computers (Cirac \& Zoller 1995).  Though NMR quantum
devices as currently conceived will probably be limited by signal-to-noise
considerations to a storage capacity of order 10 qubits, these pioneering
experiments may still prove illuminating.  But to fulfill their potential, both
the NMR experiments and the ion-trap experiments should progress beyond the
stage of mere demonstrations.  A suitable goal for the NMR program would be to
probe in unprecedented detail (via quantum tomography) the mechanisms of
decoherence for the nuclear spins, with particular emphasis on quantifying the
multi-spin correlations.  It would be exciting if this program were to evolve
into a new tool for studying molecular structure.

Actually, NMR quantum computing is not really a new phenomenon -- for some
years, quantum circuits have been routinely implemented in multipulse NMR
routines.  But the quantum computing paradigm provides a powerful and
systematic new perspective on NMR techniques, and can be expected to lead to
the design of new pulse sequences for a variety of purposes.

More broadly, the emerging paradigm of quantum computing will continue to
influence experimental physics by suggesting new kinds of measurements and
experiments.  This trend is already apparent in the studies of decoherence of
entangled states reported at this meeting by Haroche (Haroche 1997; Brune {\it
et al.} 1996) and Wineland (Wineland {\it et al.} 1997; Meekhof {\it et al.}
1996). Thinking in terms of a quantum gate array broadens our perspective on
how quantum states can be manipulated and monitored (D'Helon \& Milburn 1997).
Long before quantum computers emerge as commercially viable computing devices,
they will be important tools in the physics laboratory.  I anticipate that as
the technology of quantum computation progresses, it will be used for high
precision studies of decoherence in quantum systems, and the insights gleaned
from these studies will in turn be incorporated into more sophisticated error
correction schemes that can enhance our ability to resist and combat
decoherence.  This program will forge an alliance between experimenters and
theorists that is bound to be highly productive, irrespective of the long term
commercial potential of quantum computation. 

Those who analyze the results of  forthcoming experiments on multiqubit
decoherence will face an interesting dilemma -- decoherence is {\it complex}.
A general superoperator (trace-preserving, completely positive linear map of
density matrices to density matrices) describing the evolution of $k$ qubits
has $4^k\left(4^k-1\right)$ real parameters; this is already 240 parameters for
just two qubits!  New ideas will be needed on how to organize the data so that
it can be given a useful and meaningful interpretation.

The advance of the frontier of experimental quantum computation should be
accompanied by a parallel advance in numerical simulation (on classical
computers) of quantum systems (Despain and Obenland 1996, 1997; Miquel {\it et
al.} 1996, 1997; Barenco {\it et al.} 1996).  A quantum circuit is a strongly
coupled system, and its scaling properties are not obvious.  Currently, precise
simulations of quantum circuits are limited to a modest number of qubits and
gates, and because of the unfavorable scaling of the resources needed to
perform quantum simulation, these limitations will not be easily overcome.  To
proceed to larger computers and longer computations, simulators will need to
adopt simplified models of the operation of the device.  These simplified
models will need to be validated by checking that they yield acceptable results
for smaller systems, where they can be compared with more exact simulations.
It is partly because of the difficulty of the simulations that experiments with
a few tens of qubits have the potential to yield surprises.\footnote{One
particularly interesting challenge for both simulation and experiment will be
the behavior of qubits in close proximity, for example trapped ions with a
separation comparable to the wavelength of visible light.  Little is currently
known about how such systems will behave.}

On the theoretical front, it is important to emphasize that the work of the
past few years has already established an enduring intellectual legacy.  A new
classification of complexity has been erected, a classification better founded
on the fundamental laws of physics than traditional complexity theory.  And the
work on quantum error correction has generated profound new insights into the
nature of decoherence and how it can be controlled.  My own view is that the
development of the theory of quantum error correction may in the long run have
broader and deeper implications than the development of quantum complexity
theory.

There are many ways that theorists working in the near term could advance the
state of the art.  Here is a brief list of some interesting open issues touched
on earlier in this talk:\footnote{There are a number of appealing theoretical
problems concerning quantum information that are not included on the list
because they appear to be of tangential relevance to quantum {\it computation}.
 Particularly notable is the problem of understanding the capacity of noisy
quantum channels for sending either quantum or classical information (Lloyd
1996; Bennett {\it et al.} 1996; Shor \& Smolin 1996; Schumacher \& Nielsen
1996; Barnum {\it et al.} 1997; Bennett {\it et al.} 1997b; Holevo 1996; Fuchs
1997).}
\begin{itemize}
\item Explore and characterize the generalizations of Grover's database search
algorithm.  (What classical algorithms will admit a $\sqrt{\rm Time}$ quantum
speedup?)
\item Sharpen the proposal to use quantum computers for quantum simulation. 
\item Seek quantum algorithms for problems that are suspected to lie in the NPI
class (such as the graph isomorphism problem). 
\item Explore the applications of quantum computers to problems outside the NP
class.
\item Understand more deeply what makes quantum algorithms work. (This insight
may illuminate the path to new algorithms.)
\item Identify universal features of the ``phase transition'' between quantum
and classical devices.
\item Characterize the general intrinsic limits on the accuracy and speed of
quantum gates.
\item Adapt the methods of fault-tolerant computing to more general error
models, and to realistic devices.
\item Seek more efficient ways to implement error recovery and fault-tolerant
quantum gates (which would weaken the accuracy requirements for reliable
computation).
\item Find broader realizations of quantum error correction (beyond the
abstract quantum circuit model).
\item Conceive new ways to use quantum computation to measure interesting
observables that are otherwise inaccessible.
\item Conceive new (commercial?) applications of small-scale quantum
computers.
\item Extend simulations of quantum computers to larger devices and longer
computations by adopting (validated) simplified models.
\item Formulate a concrete program applying NMR and ion-trap computing to the
study of multi-qubit decoherence.
\item Find new ways to organize and interpret experimental data pertaining to
multi-qubit decoherence.
\item Think of good questions that are not on this list.
\end{itemize}

Surveying this list of challenges reminds us that the development of quantum
computation will require the efforts of people with expertise in a wide variety
of disciplines, including mathematics, computer science and information theory,
theoretical and experimental physics, chemistry, and engineering.  This
interdisciplinary character is one of the most exhilarating and appealing
aspects of quantum computation.

\bigskip

Serge Haroche, while a leader at the frontier of experimental quantum
computing, continues to deride the vision of practical quantum computers as an
impossible dream that can come to fruition only in the wake of some as yet
unglimpsed revolution in physics (Haroche 1997). As everyone at this meeting
knows well, building a quantum computer will be an enormous technical
challenge, and perhaps the naysayers will be vindicated in the end.  Surely,
their skepticism is reasonable.  But to me, quantum computing is not an
impossible dream; it is a possible dream.  It is a dream that can be held
without flouting the laws of physics as currently understood. It is a dream
that can stimulate an enormously productive collaboration of experimenters and
theorists seeking deep insights into the nature of decoherence. It is a dream
that can be pursued by responsible scientists determined to explore, without
prejudice, the potential of a fascinating and powerful new idea. It is a dream
that could change the world. So let us dream.

\bigskip

This work has been supported in part by the Department of Energy under Grant
No. DE-FG03-92-ER40701, and by DARPA under Grant No. DAAH04-96-1-0386
administered by the Army Research Office. I thank David DiVincenzo and Wojciech
Zurek for organizing this stimulating meeting, and for giving me this
opportunity to express my views. My thinking about quantum computing has been
influenced by discussions with many people, including Dave Beckman, Al Despain,
Eddie Farhi, Jeff  Kimble, Alesha Kitaev, Manny Knill, Raymond Laflamme, Seth
Lloyd, and Peter Shor.  I am particularly grateful to Gilles
Brassard, Ike Chuang, David DiVincenzo, Chris Fuchs, Rolf Landauer, Hideo
Mabuchi, Martin Plenio, Dave Wineland, and Christof Zalka for helpful comments on the
manuscript.  I especially wish to thank Michael Nielsen for many detailed
suggestions, and Daniel Gottesman for countless discussions of all aspects of
quantum computation.

\bigskip

\leftline{\Large \bf References}
\begin{description}
\item Abrams, D.~S. \& Lloyd, S. 1997 Simulation of many-body fermi systems on
a universal quantum computer. (Online preprint quant-ph/9703054.)
\item Aharonov, D. \& Ben-Or, M. 1996a Fault tolerant quantum computation with
constant error. (Online preprint quant-ph/9611025.)
\item Aharonov, D. \& Ben-Or, M. 1996b Polynomial simulations of decohered
quantum computers. (Online preprint quant-ph/9611029.)
\item Barenco, A., Brun, T.~.A., Schack, R. \& Spiller, T. 1996 Effects of
noise on quantum error correction algorithms. (Online preprint
quant-ph/9612047.)
\item Barnum, H., Nielsen, M.~A. \& Schumacher, B. 1997 Information
transmission through a noisy quantum channel. (Online preprint
quant-ph/9702049.)
\item Benioff, P. 1982 Quantum mechanical models of Turing machines that
dissipate no energy. {\it Phys. Rev. Lett.} {\bf 48}, 1581.
\item Bennett, C.~B., Bernstein, E., Brassard, G. \&
Vazirani, U. 1997a Strengths and weaknesses of quantum computing.
(Online preprint quant-ph/9701001.)
\item Bennett, C.~H. \& Brassard, G. 1984. In {\it Proceedings of IEEE
International Conference on Computers, Systems, and Signal Processing},
Bangalore, India. New York: IEEE, p. 175.
\item Bennett, C., DiVincenzo, D., Smolin, J. \& Wootters, W. 1996
Mixed state entanglement and quantum error correction. {\it Phys. Rev.} A
{\bf 54}, 3824.
\item Bennett, C.~H., DiVincenzo, D.~P. \& Smolin, J.~A. 1997b
Capacities of quantum erasure channels. (Online preprint quant-ph/9701015).
\item Bernstein, E. \& Vazirani, U. 1993 Quantum complexity theory. In {\it
Proceedings of the 25th ACM Symposium on the Theory of Computation}.  New York:
ACM Press, pp. 11-20.
\item Boghosian, B.~M. \& Taylor, W. 1997 Simulating quantum mechanics on a
quantum computer.  (Online preprint quant-ph/9701019.)
\item Bollinger, J.~J., Itano, W.~M., Wineland, D.~J. \& Heinzen, D.~J. 1997
Optical frequency measurements with maximally correlated states.  {\it Phys.
Rev.} A {\bf 54}, R4649.
\item Braginsky, V.~B., Khalili, F.~ya. \& Sazhin, M.~V. 1995 Decoherence in
e.m. vacuum. {\it Phys. Lett.} A {\bf 208}, 177.
\item Brune, M., Hagley, E., Dreyer, J., Maitre, X., Maali, A., Wunerlich, C.,
Raimond, J.~M. \& Haroche, S. 1996 Observing the progressive decoherence of the
meter in a quantum measurement.  {\it Phys. Rev. Lett.} {\bf 77}, 4887.
\item Calderbank, A.~R. \& Shor, P.~W. 1996 Good quantum
error-correcting codes exist. {\it Phys. Rev.} A {\bf 54}, 1098.
\item  Cirac, J.~I. \& Zoller, P. 1995 Quantum computations
with cold trapped ions. {\it Phys. Rev. Lett.} {\bf 74}, 4091.
\item Cirac, J.~I., Pellizzari \& Zoller, P. 1996 Enforcing coherent evolution
in dissipative quantum dynamics. {\it Science} {\bf 273}, 1207.
\item Cirac, J.~I., Zoller, P., Kimble, H.~J. \& Mabuchi, H. 1997  Quantum
state transfer and entanglement distribution among distant nodes in a quantum
network. {\it Phys. Rev. Lett.} {\bf 78}, 3221.
\item Cory, D.~G., Fahmy, A.~F. \& Havel, T.~F. 1996 Nuclear magnetic resonance
spectroscopy: an experimentally accesible paradigm for quantum computing. In
{\it Proceedings of the 4th Workshop on Physics and Computation}, Boston: New
England Complex Systems Institute.
\item Deutsch, D. 1985 Quantum theory, the Church-Turing principle and the
universal quantum computer. {\it Proc. Roy. Soc. Lond.} A {\bf 400}, 96. 
\item D'Helon, C. \& Milburn, G.~J. 1997 Quantum measurements with a quantum
computer. (Online preprint quant-ph/9705014.)
\item DiVincenzo, D. \& Shor, P. 1996 Fault-tolerant error
correction with efficient quantum codes. {\it Phys. Rev. Lett.} {\bf 77},
3260.
\item Feynman, R.~P. 1982 Simulating physics with computers.
{\it Int. J. Theor. Phys.} {\bf 21}, 467.
\item Fuchs, C. 1997 Nonorthogonal quantum states maximize classical
information capacity. (Online preprint quant-ph/9703043.)
\item Garey, M.~R. \& Johnson, D.~S. 1979 {\it Computers and intractability: a
guide to the theory of NP-completeness}. New York: W.~H. Freeman and Co.
\item Garg, A. 1996 Decoherence in ion-trap quantum computers. {\it Czech. J.
Phys.} {\bf 46}, 2375.
\item Gershenfeld, N. \& Chuang, I. 1997 Bulk spin resonance
quantum computation. {\it Science} {\bf 275}, 350.
\item Gottesman, D., Evslin, J., Kakade, S. \& Preskill, J. 1996, to be
published.
\item Gottesman, D. 1997 A theory of fault-tolerant
quantum computation. (Online preprint quant-ph/9702029.)
\item Grassl, M., Beth, Th. \& Pellizzari, T. 1996 Codes for the
quantum erasure channel. (Online preprint quant-ph/9610042.)
\item Grover, L.~K. 1996 A fast quantum mechanical algorithm for
database search. {\it Proceedings, 28th ACM Symposium on Theory of
Computation},
212.
\item Haroche, S. 1997, these proceedings.
\item Haroche, S. \& Raimond, J.~M. 1996 Quantum computing: dream or nightmare?
{\it Phys. Today} {\bf 49} (8), 51.
\item Huelga, S.~F., Macchiavello, C., Pellizzari, T., Ekert, A.~K., Plenio, M.~B., \& Cirac, J.~I. 1997 On the improvement of frequency standards with quantum entanglement. (Online preprint quant-ph/9707014.) 
\item Holevo, A.~S. 1996 The capacity of quantum channel with general signal
states.  (Online preprint quant-ph/9611023.)
\item Kitaev, A.~Yu. 1996a Quantum error correction with imperfect gates, preprint.
\item Kitaev, A.~Yu. 1996b Quantum computing: algorithms and error correction, preprint (in Russian).
\item Kitaev, A.~Yu. 1997 Fault-tolerant quantum computation by anyons, to be
published.
\item Knill, E. \& Laflamme, R. 1996 Concatenated
quantum codes. (Online preprint quant-ph/9608012.)
\item Knill, E., Laflamme, R. \& Zurek, W. 1996 Accuracy
threshold for quantum computation. (Online preprint quant-ph/9610011.)
\item Knill, E., Laflamme, R. \&  Zurek, W. 1997 Resilient quantum computation:
error models
and thresholds. (Online preprint quant-ph/9702058.)
\item Laflamme, R., Miquel, C., Paz, J.~P. \& Zurek, W. 1996
Pefect quantum error correction code. {\it Phys. Rev. Lett.} {\bf 77}, 198.
\item Landauer, R. 1995 Is quantum mechanics useful? {\it Phil. Tran. R. Soc.
Lond.} {\bf 353}, 367.
\item Landauer, R. 1996 The physical nature of information.  {\it Phys. Lett.}
A {\bf 217}, 188.
\item Landauer, R. 1997 Is quantum mechanically coherent computation useful? In
{\it Proc. Drexel-4 Symposium on Quantum Nonintegrability-Quantum-Classical
Correspondence}, Philadelphia, PA, 8 September 1994 (ed. D.~H. Feng and B.-L.
Hu), Boston: International Press.
\item Leggett, A.~J., Chakravarty, S., Dorsey, A.~T., Fisher, M.~P.~A., Garg,
A. \& Zwerger, W. 1987 Dynamics of the dissipative two-state system.  {\it Rev.
Mod. Phys.} {\bf 59}, 1.
\item Lidar, D.~A. \& Biham, O. 1996 Simulating Ising spin glasses on a quantum computer. (Online preprint quant-ph/9611038.)
\item Lloyd, S. 1993 A potentially realizable quantum computer.  {\it Science}
{\bf 261}, 1569.
\item Lloyd, S. 1996 Universal quantum simulators. {\it Science} {\bf 273},
1073.
\item Lloyd, S. 1997 The capacity of a noisy quantum channel.
{\it Phys. Rev.} A {\bf 55}, 1613.
\item Mabuchi, H. and Zoller, P. 1996 Inversion of quantum jumps in
quantum-optical systems under continuous observation. {\it Phys. Rev. Lett.}
{\bf 76}, 3108.
\item Meekhof, D.~M., Monroe, C., King, B.~E., Itano, W.~M. \& Wineland, D.~J.
1996 Generation of nonclassical motional states of a trapped atom.  {\it Phys.
Rev. Lett.} {\bf 76}, 1796.
\item Meyer, D.~A. 1996 Quantum mechanics of lattice gas automata I: one
particle plane waves and potentials.  (Online preprint quant-ph/9611005.)
\item Miquel, C., Paz, J.~P. \& Perazzzo, R. 1996 Factoring in a dissipative
quantum computer. {\it Phys. Rev.} A {\bf 54}, 2605.
\item Miquel, C., Paz, J.~P. \& Zurek, W.~H. 1997 Quantum computation with phase
drift errors. (Online preprint quant-ph/9704003.)
\item Monroe, C., Meekhof, D.~M., King, B.~E., Itano, W.~M. \&
Wineland, D.~J. 1995 Demonstration of a fundamental quantum logic gate. {\it
Phys. 
Rev. Lett.} {\bf 75}, 4714.
\item Obenland, K. \& Despain, A.~M. 1996a Simulation of factoring on a quantum
computer architecture. In {\it Proceedings of the 4th Workshop on Physics and
Computation}, Boston, November 22-24, 1996, Boston: New England Complex Systems
Institute.
\item Obenland, K. \& Despain, A.~M. 1996b Impact of errors on a quantum
computer architecture. (Online preprint {\tt
http://www.isi.eu/acal/quantum/quantum\_op\_errors.ps}.)
\item Pellizzari, T., Gardiner, S.~A., Cirac, J.~I. \& Zoller, P. 1995
Decoherence, continuous observation, and quantum computing: a cavity QED
model.
{\it Phys. Rev. Lett.}  {\bf 75}, 3788.
\item Plenio, M.~B.  \& Knight, P.~L. 1996 Decoherence limits to quantum
computation using trapped ions. (Online preprint quant-ph/9610015.)
\item Preskill, J. 1997 Reliable quantum computers. (Online preprint
quant-ph/9705031.)
\item Schumacher, B. \& Nielsen, M.~A. 1996 Quantum data
processing and error correction. {\it Phys. Rev.} A {\bf 54}, 2629.
\item Shor, P. 1994 Algorithms for quantum computation:
discrete logarithms and factoring. In {\it Proceedings of the 35th Annual
Symposium
on Fundamentals of Computer Science}. Los Alamitos, CA: IEEE Press, pp.
124-134.
\item Shor, P. 1995 Scheme for reducing decoherence in quantum
memory. {\it Phys. Rev.} A {\bf 52}, 2493.
\item Shor, P. 1996 Fault-tolerant quantum computation. In {\it Proceedings of
the Symposium on the Foundations of Computer Science}. Los Alamitos, CA: IEEE
Press (Online preprint  quant-ph/9605011).
\item Shor, P. \& Smolin, J. 1996 Quantum error-correcting codes
need not completely reveal the error syndrome. (Online preprint
quant-ph/9604006.)
\item Simon, D.~R. 1994 On the power of quantum computation. In {\it
Proceedings of the 35th Annual Symposium
on Fundamentals of Computer Science}. Los Alamitos, CA: IEEE Press, pp.
116-123.
\item Steane, A.~M. 1996a Error correcting codes in quantum
theory. {\it Phys. Rev. Lett.} {\bf 77}, 793.
\item Steane, A.~M. 1996b Multiple particle interference and quantum
error correction. {\it Proc. Roy. Soc. Lond.} A {\bf 452}, 2551.
\item Steane, A.~M. 1997 Active stabilization, quantum
computation and quantum state synthesis. {\it Phys. Rev. Lett.} {\bf 78},
2252.
\item Turchette, Q.~A., Hood, C.~J., Lange, W., Mabuchi, H. \&
Kimble, H.~J. 1995 Measurement of conditional phase shifts for quantum logic. 
{\it Phys. Rev. Lett.} {\bf 75}, 4710.
\item Unruh, W.~G. 1995 Maintaining coherence in quantum computers. {\it Phys.
Rev.} A {\bf 51}, 992.
\item Van Enk, S.~J., Cirac, J.~I. \& Zoller, P. 1997 Quantum communication
over noisy channels: a quantum optical implementation. (Online preprint
quant-ph/9702036.)
\item Wiesner, S. 1996 Simulations of many-body quantum systems by a quantum
computer. (Online preprint quant-ph/9603028.)
\item Wineland, D.~J., Monroe, C., Meekhof, D.~M., King, B.~E., Liebfried, D.,
Itano, W.~M., Bergquist, J.~C., Berkeland, D., Bollinger, J.~J. \& Miller, J.
1997 Quantum state manipulation of trapped atomic ions. (Online preprint
quant-ph/9705022.)
\item Wootters, W.~K. \& Zurek, W.~H. 1982 A single quantum
cannot be cloned. {\it Nature} {\bf 299}, 802.
\item Zalka, C. 1996a Efficient simulation of quantum systems by quantum computers. (Online preprint quant-ph/9603026.)
\item Zalka, C. 1996b Threshold estimate for fault tolerant quantum
computing. (Online preprint quant-ph/9612028.)

\end{description}
\end{document}